\documentclass[useAMS,usenatbib,usegraphicx]{mn2e}

\newcommand{\gap}{\;\rlap{\lower 2.5pt \hbox{$\sim$}}\raise 1.5pt\hbox{$>$}\;}
\newcommand{\lap}{\;\rlap{\lower 2.5pt \hbox{$\sim$}}\raise 1.5pt\hbox{$<$}\;}
\newcommand{\beq}{\begin{equation}}
\newcommand{\eeq}{\end{equation}}

\title[The nature of the HE0450-2958 system]{The nature of the HE0450-2958 system}
\author[D. Merritt et al.]
{David Merritt$^{1}$,
Thaisa Storchi Bergmann$^{2}$,
Andrew Robinson$^{1}$,
Dan Batcheldor$^{1}$,
\newauthor
David Axon$^{1}$, and
Roberto Cid Fernandes$^{3}$
\\
$^{1}$Department of Physics, Rochester Institute of Technology, 84 Lomb Memorial Drive, Rochester, NY 14623\\
$^{2}$ Instituto de F\'isica, UFRGS, Porto Alegre, RS, Brazil\\
$^{3}$Departamento de F\'isica, CFM-UFSC, Florian\'oplis, SC, Brazil}

\begin{document}

\maketitle

\label{firstpage}

\begin{abstract}
Magain et al. (2005) argued that the host galaxy of the
quasar in HE0450-2958 is substantially under-luminous
given the likely mass of its nuclear black hole.
Using kinematical information from the spectra of the
quasar and the companion galaxy, an ultra-luminous
infrared galaxy, we test the hypothesis that the
black hole powering the quasar was ejected from the 
companion galaxy during a merger.
We find that the ejection model can be securely ruled out,
since the kick velocity required to remove the black hole 
from the galaxy is $\gap 500$ km/s, 
inconsistent with the presence of narrow emission line
gas at the same redshift as the quasar nucleus.
We also show that the quasar in HE0450-2958 has the spectral
characteristics of a narrow-line Seyfert 1 galaxy and calculate a mass for 
its black hole that is roughly an order of magnitude 
smaller than estimated by Magain et al.
The predicted luminosity of the host galaxy is then
consistent with the upper limits inferred by those authors.
\end{abstract}

\section{Introduction}

HE0450-2958 is a bright quasar at redshift $z=0.285$.
HST images revealed that the system is double,
with an ultra-luminous infrared galaxy (ULIRG)
situated $\sim 1.5$ arcsec from the quasar 
\citep{boyce-96,canalizo-01}.
Recently, \cite{magain-05} reported that the host galaxy
of the quasar is substantially under-luminous, based on 
the quasar's luminosity and on a likely value for $M_\bullet$,
the mass of its nuclear supermassive black hole (SBH).
Magain et al. proposed either that the quasar host galaxy is dark,
or that an otherwise ``naked'' SBH had acquired gas while
moving through intergalactic space.

Here we examine these hypotheses in light of additional 
evidence from the spectra of the quasar and the companion galaxy.
The quasar spectrum reveals it to be a
typical narrow-line Seyfert 1 galaxy \citep{osterbrock-85},
not a giant elliptical galaxy as assumed by \cite{magain-05}.
We infer a much smaller luminosity for the host
galaxy, consistent with the upper limits derived
by those authors.
We also critically examine the most natural model for 
a ``naked'' SBH, namely, a SBH that was ejected from
the companion galaxy during the merger that created the ULIRG
\citep{merritt-04}.
We show that the ejection model can be securely ruled out,
since the quasar spectrum indicates the presence
of narrow emission line gas extending out to a distance of
$\sim 1$ kpc from the nucleus that is moving at the same velocity 
as the broad-line gas.
The narrow-line gas could not have remained bound to the SBH
if it were ejected from the companion  galaxy.

\section{Spectral Analysis}

HE0450-2958 was observed during November 27 2001 using the UV
Focal Reducer and low-dispersion spectrograph (FORS1) on Unit Telescope 1
of the VLT (PI: M. Courbin). 
The instrument was operated in MOS mode with a
long-slit position angle of $\sim55\degr$. This allowed the contributions
from the ULIRG, quasar and the nearby G-type star to be gathered simultaneously
in slit no. 9. 
In total, five spectra were obtained of HE0450-2958: 
three of 1200 s duration using the 600B grism centered at 4620\AA, 
and two of 1800 s duration using the 600R (6270\AA) and 600I (7940\AA) grisms. 
These data, including all relevant calibration files, 
were retrieved from the VLT data archive
\footnote{\tt http://archive.eso.org/}.

The data were reduced using standard IRAF\footnote{IRAF is distributed
by the National Optical Astronomy Observatories, which are operated by the
Association of Universities for Research in Astronomy, Inc., under
cooperative agreement with the National Science Foundation.} routines. Bias
and flat-field subtraction were carried out before wavelength calibration
with HeArNe arcs. Cosmic ray subtraction was facilitated with a median
combine, in the case of the three 600B exposures, and with a median filter
and rigorous visual inspection in the 600R and 600I cases. Background
subtraction was performed by removing a third-order polynomial fitted to the
sky components in the spatial direction. Flux calibration was carried out by
fitting a 5700K (G-type) black body spectrum to the observed star, and
scaling to the fluxes observed through the High Resolution Channel of the
Advanced Camera for Surveys aboard the {\it Hubble Space Telescope}
(no. 10238, PI: Courbin).

\begin{figure}
\includegraphics[width=8.5cm]{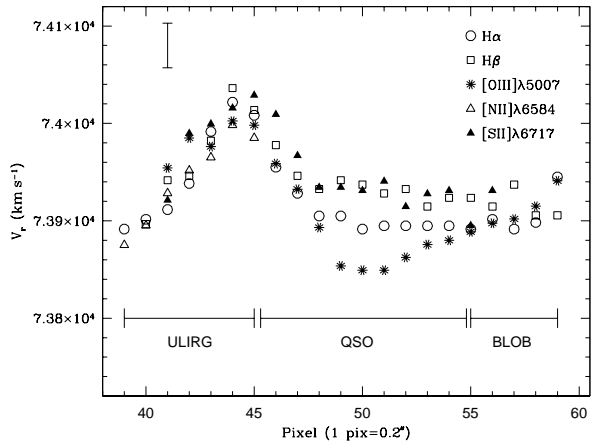}
\caption{
Radial velocities obtained from the peak wavelengths of the strongest
emission lines along the slit. 
The locations of the galaxy, quasar and ``blob'' identified by 
Magain et al. (2005) are indicated. }
\label{fig:kin}
\end{figure}

Figure~\ref{fig:kin} shows the velocities derived from the
peak wavelengths of the strongest emission lines along the slit; 
the regions dominated by the ULIRG, 
the quasar and the \cite{magain-05} ``blob'' are indicated.
In the region dominated by the galaxy, 
the velocities show the pattern of a rotation curve, 
with peak-to-peasuch a k amplitude of $\sim140$ km s$^{-1}$. 
This value seems small for such a luminous galaxy, but the HST image
suggests that the orientation of the slit, aligned to include the
star, quasar and ULIRG, is far from the major axis of the galaxy.
According to the rotation curve, the systemic velocity of the 
galaxy is 73950$\pm$20\,km\,s$^{-1}$. 
In the region dominated by the quasar, the velocities obtained from the
H$\alpha$, H$\beta$, [NII]$\lambda$6584 and [SII]$\lambda$6717
emission lines do not vary within the uncertainties, and the average
value is 73920$\pm$20\,km\,s$^{-1}$, indicating a blueshift
relative to the systemic velocity of the galaxy of only 30\,km\,s$^{-1}$,
consistent within the errors with zero.
The [OIII]$\lambda$5007 emission line is blue-shifted relative
to the other emission lines by approximately 60\,km\,s$^{-1}$.
However such a blueshift in [OIII]$\lambda$5007 is often observed in 
AGN \citep{nelson-95,bian-05,boroson-05}.
In the region dominated by the emission of the blob,
this blueshift is not observed, and the velocities
derived from the [OIII] emission line are the same as those
obtained from the other lines. 
In summary, our measurements show
similar systemic velocities for the ULIRG, quasar and blob.

\begin{figure}
\includegraphics[width=8.cm]{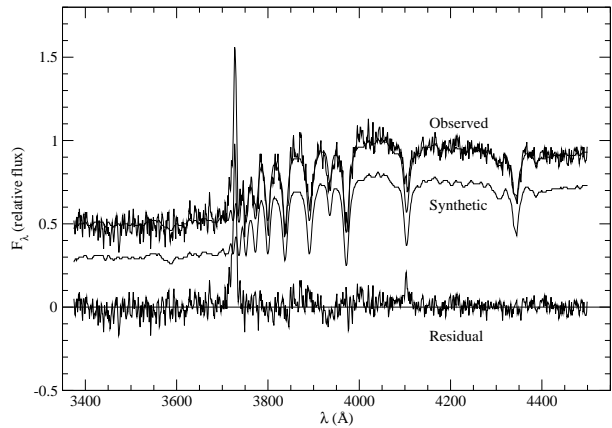}
\caption{
The blue spectrum of the ULIRG, together with the synthetic
spectrum and the residual between the two. The synthetic spectrum
is ploted twice: superimposed on the galaxy spectrum and shifted
vertically for clarity.}
\label{fig:synthesis}
\end{figure}

The blue spectrum of the galaxy (Figure~\ref{fig:synthesis})
shows a number of absorption-line features,
in particular, high-order Balmer lines indicative
of an intermediate-age ($\sim\,10^8$yr) stellar population.
In order to better quantify the age of the stellar population 
and to obtain an estimate for the stellar velocity dispersion, 
we performed a spectral synthesis using the code STARLIGHT of 
\cite{fernandes-05}. 
This code uses a basis of stellar population templates, 
each one corresponding to a given metallicity and age,
to synthesize the galaxy spectrum, and it gives as output the
contribution of each template to the total light at $\lambda$4020\AA;
the internal reddening; and the velocity dispersion.
The contribution of the quasar, which contaminates both
the continuum and the emission lines of the galaxy spectrum, 
has also been included in the synthesis. 
The results of the synthesis (Figure 2) yield the following
contributions to the light of the ULIRG at $\lambda$4020: 
 $\sim$25\% from the quasar continuum, 
$\sim$50\% from stars with ages around $10^8$\,yr
and $\sim$25\% from older stars.
The total stellar mass is $\sim 6\times 10^{10}M_\odot$,
with the major uncertainty due to the quasar continuum contribution.
The internal reddening is A$_V$=0.4 mag and the velocity dispersion is 
$\sigma=190\pm\,25\,$ km\,s$^{-1}$.
The synthesis strongly points to a major burst
of star formation $\sim 10^8$\,yr ago.

\begin{table*}
\caption{Mass Models \label{tab:1}}
\begin{center}
\begin{tabular}{c|ccc|ccc|c}
\hline
        &     & galaxy &           &             &halo       &           & both      \\
        & $M$ & $R_e$  & $V_{esc}$ & $M$         & $r_{1/2}$  & $V_{esc}$ & $V_{esc}$ \\ 
        & ($10^{11} M_\odot$) & (kpc) & (km s$^{-1}$) & ($10^{12} M_\odot$) & (kpc) & (km s$^{-1}$) & (km s$^{-1}$) \\
& & & & & & & \\
Model 1 & $0.15$ & $1.10$ & $440$  & $2.0$ & $200$ & $500$ & $664$ \\
Model 2 & $1.50$ & $1.43$ & $1520$ & $2.0$ & $200$ & $500$ & $1597$ \\
\hline
\end{tabular}
\end{center}
\end{table*}

\section{Ejection Hypothesis}

The low luminosity of the quasar host, coupled with
its proximity to a ULIRG, leads naturally to the 
hypothesis that the SBH powering the
quasar was ejected from the ULIRG following a merger.
Two ejection mechanisms have been discussed:
gravitational radiation recoil during
the coalescence of a binary SBH \citep{favata-04};
or a gravitational slingshot involving three SBHs,
if the merger happened to bring a third SBH into the
center of a galaxy containing an uncoalesced binary
\citep{mikkola-90}.

The quasar is displaced $1.5$ arcsec from the center of the ULIRG,
corresponding to a projected separation of $\sim 6.5$ kpc.
This is much greater than the galaxy's half-light radius
implying an ejection velocity $V_{kick}$ comparable to the 
central escape velocity from the galaxy $V_{esc}$.
The distribution of mass around the companion galaxy of HE0450-2958
is unknown.
However, \cite{tacconi-02} find that
the light distributions in a sample of 18 ULIRGs are
reasonably well fit by de Vaucouleurs profiles with effective
(projected half-light) radii of $R_e\approx 1$ kpc, similar to
those of luminous E galaxies.
They derive kinematical masses of 
$0.3\times 10^{11}M_\odot\lap M\lap 5\times 10^{11}M_\odot$,
consistent with the estimate presented above from population
synthesis of $0.5-0.8\times 10^{11}M_\odot$.
The mean stellar velocity dispersion in their sample is $180$ km s$^{-1}$,
consistent with the population synthesis estimates presented above.
We accordingly modelled the baryonic mass distribution around the 
ULIRG as if it were a normal, spherical E galaxy of mass $M_{gal}$, 
and used the empirical correlations between E-galaxy mass, luminosity,
effective radius and Sersic index 
\citep{magorrian-98,gg-03,acs6}
to derive its gravitational potential.

Table ~\ref{tab:1} gives the parameters of two mass models 
for the companion galaxy.
Model 1 has a baryonic mass of $1.5\times 10^{10}M_\odot$,
roughly the mass of a $M_B\approx -18$ dE galaxy
and a factor of $\sim$two {\it smaller} than the smallest 
ULIRG mass inferred by Tacconi et al.
Model 2 has $M_{gal} = 1.5\times 10^{11}M_\odot$, 
the stellar mass of a $M_B\approx -20$ E galaxy, 
and close to the average mass of the galaxies in the Tacconi et al. sample.
We also included a dark-matter halo; because the contribution
of dark matter to the gravitational force on scales $\lap 10$ kpc
is probably much less than that of the baryons,
we considered only a single halo model.
As templates for the dark matter, we considered the four
``galaxy-sized'' halos in the \cite{diemand-04} $\Lambda$CDM 
simulations, which have virial masses in the range 
$1.0\times 10^{12}M_\odot \le M_{DM} \le 2.2\times 10^{12}M_\odot$.
Fits to $\rho(r)$ for these halos are given in \cite{graham-05};
based on these fits, central escape velocities lie in the range
$480\ {\rm km\ s}^{-1} \le V_{esc} \le 600\ {\rm km\ s}^{-1}$, and the
$\Delta V$ in climbing to $10$ kpc is
$210\ {\rm km\ s}^{-1} \le \Delta V \le 310$ km s$^{-1}$.
Our adopted halo model (Table~\ref{tab:1}) had a mass of
$2.0\times 10^{12}M_\odot$ (virial mass $1.1\times 10^{12}M_\odot$), 
half-mass radius $200$ kpc and
central escape velocity $\sim 500$ km s$^{-1}$, similar to model G1
in \cite{diemand-04}.
By comparison, the virial mass of the Milky Way halo is believed to
be $1-2\times 10^{12}M_\odot$ \citep{klypin-02}.

In de Vacouleurs-like mass models, the $\Delta V$ in climbing from the
very center out to a distance of a few parsecs can be considerable
due to the high nuclear density (e.g. \cite{young-76}).
The mass distribution of the companion galaxy following the
merger is unknown on these small radial scales, 
and in any case, the SBH would carry with it
the mass of the inner few parsecs, modifying the potential.
Accordingly, we placed the SBH initially at a distance of
$10$ pc from  the  galaxy center when computing post-ejection
trajectories.

\begin{figure}
\includegraphics[width=8.5cm]{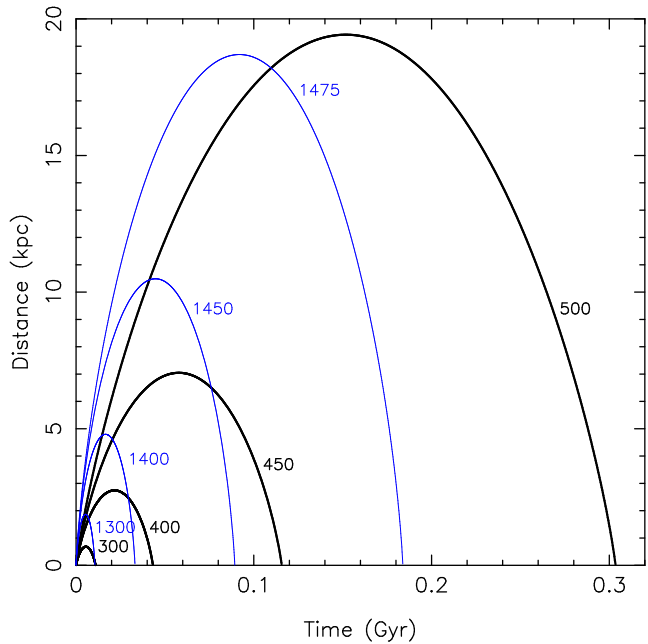}
\caption{
Trajectory of a kicked SBH in two models for the
mass distribution of the companion galaxy (Table~\ref{tab:1}).
{\it Black (thick) curves:} Model 1;
{\it blue (thin) curves:} Model 2.
Curves are labelled by $V_{kick}$ in km s$^{-1}$.}
\label{fig:roft}
\end{figure}

Figure~\ref{fig:roft} shows the results.
As expected, kicks of $\sim 500$ km s$^{-1}$ (Model 1)
and $\sim 1500$ km s$^{-1}$ (Model 2) are required in order for the
SBH to climb a distance of $10$ kpc from the center of the galaxy.
{\it This result essentially rules out radiation recoil as
the origin of the kick},
since the maximum amplitude of the recoil is believed
to be less than $500$ km s$^{-1}$ \citep{merritt-04}
and probably no more than $250$ km s$^{-1}$ \citep{blanchet-05}.
Three-body recoils might still work, however
the largest $\Delta V$ in a three-body interaction
is experienced by the {\it smallest} body.
But Figure~\ref{fig:roft} constrains {\it any} 
ejection model in a number of other ways:

1. A large $V_{kick}$ implies a large velocity,
$V\gap 300$ km s$^{-1}$, as the SBH moves past its current position.
This is hard to reconcile with the essentially zero
radial velocity difference between quasar and galaxy,
unless the ejection velocity is fine-tuned or nearly perpendicular
to the line of sight.

2. The time for the kicked SBH to reach a distance of $10$ kpc 
is much shorter than $10^8$ yr, again unless the gravitational
potential and kick velocity are finely tuned.
But the starburst occurred $\sim 10^8$ yr ago; 
thus, either the true separation of the SBH from the ULIRG
is much greater than $10$ kpc, or the ejection was delayed 
until a time of $\sim 10^8$ yr after the starburst.

3. A large $V_{kick}$ implies that the ejected SBH will carry 
very little mass with it as it departs the galaxy.
Material orbiting the SBH with velocity
$v\gg V_{kick}$ before the kick
will experience the kick as an adiabatic perturbation
and will ``instantaneously'' acquire
the specific momentum of the SBH.
This argument suggests that the SBH will carry with
it the mass contained initially within a region 
whose size is less than $r_{eff}$, the radius at which
the orbital velocity around the SBH is equal to $V_{kick}$,
or
\begin{equation}
r_{eff} = {GM_\bullet\over \sigma^2}
\left({V_{kick}\over\sigma}\right)^{-2} 
\approx 10 M_8 \sigma^{-2}_{200}
\left({V_{kick}\over\sigma}\right)^{-2}\ {\rm pc}
\label{eq:reff}
\end{equation}
with $M_8\equiv M_\bullet/10^8 M_\odot$ and
$\sigma_{200}\equiv \sigma/200$ km s$^{-1}$.
Since $M_8\approx 1$ (see below) and
$V_{kick}\gap 3 \sigma$ (Figure~\ref{fig:roft}), 
the entrained region will be of order 1 pc or less in size.
This is probably sufficient to include the 
broad-line region (BLR) gas, which is expected to have a size
$\sim 0.3$ pc based on the empirical scaling relation
between BLR size and $5007$\AA\, luminosity \citep{kaspi-05,greene-05}, 
but not larger structures.

\section{Size of the Narrow-Line Region}

The radius of the narrow-line region can be 
estimated from the ionization parameter and density of the 
emitting gas and the ionizing luminosity of the QSO. 
For an ionizing photon luminosity, $Q$, the ionization 
parameter in gas of density $n$ at a distance $r$ from the 
source can be defined as $U=Q/(4\pi R_{NLR}^2 n c)$. 
We estimate the ionizing photon luminosity by extrapolating 
the slope of the far UV ($\sim 600 - 1200$\AA) continuum as 
determined by \cite{scott-04} from {\sl FUSE} observations. 
After corrections for interstellar extinction and absorption, 
Scott et al.'s power-law ($f_\nu \propto \nu^{-\alpha}$) fit 
yields a spectral index $\alpha = -1.2\pm 0.1$ and a flux at 
1000\AA\, of $5.25\times10^{-27}$\,erg\,s$^{-1}$\,cm$^{-2}$ Hz$^{-1}$.  
HE0450-2958 has a steep soft X-ray continuum, with a photon index  
$\Gamma = 3.1$ in the 0.1-2 keV ROSAT band \citep{brinkmann-97}.
Therefore, we extrapolate the UV power-law to a high energy 
cut-off of 0.1 keV. 
Adopting a luminosity distance of 1458 Mpc (assuming 
$H_0 = 71$ km\,s$^{-1}$\,Mpc$^{-1}$ and a matter density parameter 
$\Omega_m = 0.27$) 
the integration yields $Q \approx 1.2\times 10^{56}$ photons s$^{-1}$. 

The gas density was obtained from the relative intensities of 
the [SII]$\lambda\lambda 6717,6731$ doublet. 
Although partially blended in the spectrum, Gaussian fits to 
the lines are well constrained and yield a ratio 
$I_{6717}/I_{6731} = 1.01\pm 0.05$. 
This corresponds to a density $n\approx 1000$\ cm$^{-3}$.

We determined the ionization parameter from the ratio of the 
[OII]$\lambda 3727$ and [OIII]$\lambda 5007$ lines 
\citep{baldwin-81}.
The intensity ratio measured from the spectrum is 
$I_{3727}/I_{5007} = 0.13\pm 0.05$. 
Using the empirical relation between this ratio and the ionization 
parameter given by \cite{penston-90}, we obtain $U\approx 0.014$. 
This value is broadly consistent with the more recent photoionization 
models presented by \cite{groves-04}.

Combining estimates for $Q$, $n$ and $U$, we arrive at a radius 
$R_{NLR} \approx 1.5$ kpc. 
This of course represents an ill-defined spatial average, 
since we have used integrated fluxes of emission lines representing 
different ionization states. 
Moreover, the determination of $U$ is model-dependent and extrapolating 
the far UV power-law is, at best, a crude representation of the EUV 
continuum.\footnote {However, the H$\alpha$ photon luminosity is 
$\approx 10^{56}$ photons s$^{-1}$ which, assuming photoionization, 
implies that our value for $Q$ is {\em underestimated} by at least 
a factor $\sim 3$.} 
Based on these considerations, $R_{NLR}$ may be uncertain by a factor 
$\sim 2$. 
It is, nevertheless, unsurprising that the NLR in a relatively 
luminous quasar extends to a few kpc. 
Direct imaging of bright, low-redshift quasars in [OIII] with 
HST reveals NLR sizes ranging from 1.5 kpc to 10 kpc \citep{bennert-02}. 
Indeed, the [OIII] luminosity of HE0450-2958 
($L_{[OIII]} \approx 3.6\times10^{43}$\,erg\,s$^{-1}$) 
makes it comparable with the most luminous object in 
Bennert et al.'s sample, for which they determine an NLR radius of 
10.5 kpc.          

{\it Such a large size for the NLR rules out the possibility that the NLR gas would remain bound to the SBH after ejection from the ULIRG} 
(cf. equation~\ref{eq:reff}).
Post-ejection accretion of the NLR gas from a cloud is also unlikely, 
since the radius of the Bondi accretion column, $r_{acc}$, is given by an 
equation similar to eq.~\ref{eq:reff}, after replacing $V_{kick}$ 
by the relative velocity between SBH and gas cloud,
implying $r_{acc}\ll 1$ kpc unless the ejected SBH has
fortuitously matched velocities with the cloud.

\section{Mass of the Black Hole and Implications for the Host Galaxy Luminosity}

\begin{figure}
\includegraphics[width=8.cm]{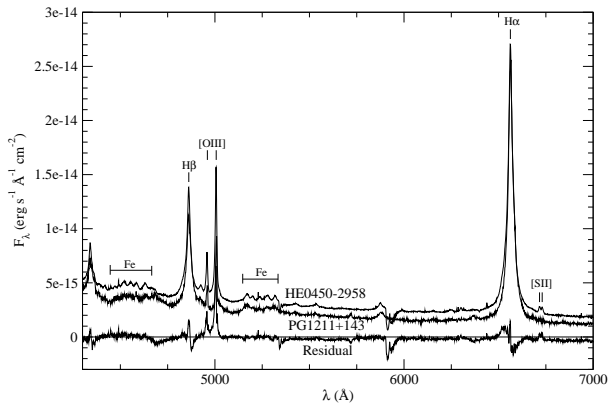}
\caption{
Comparing the quasar spectrum with that of 
PG1211+143, a radio-quiet quasar with steep $X$-ray
spectral index that is classified spectroscopically
as a NLS1 \citep{constantin-03}.
HE0450-2958 is plotted in absolute flux units,
PG1211+143 in absolute units minus 
${\rm 10^{-15} erg s^{-1} \AA^{-1} cm^{-2}}$.
Residuals are plotted along the bottom with a solid straight 
line highlighting zero.}
\label{fig:nls1}
\end{figure}

As shown in Figure~\ref{fig:nls1},
{\it HE0450-2958 exhibits characteristics which unambiguously identify 
it as a narrow-line Seyfert 1} (NLS1; \cite{grupe-04}). 
Specifically, its broad Balmer lines have FWHM's $\approx 1300$\,km\,s$^{-1}$
(the conventional definition requires FWHM $< 2000$\,km\,s$^{-1}$; 
\cite{osterbrock-85}), 
it has strong optical Fe\,II emission features and, 
as already noted, it has a steep soft X-ray photon continuum. 
The currently-favored picture of NLS1's is that they represent an 
extreme AGN population characterized by relatively low-mass 
SBHs but high accretion rates \citep{peterson-00,boroson-02} 
--- possibly substantially super-Eddington \citep{boller-05}.
It follows that estimating $M_\bullet$ from the quasar 
luminosity assuming a sub-Eddington accretion rate, 
as was done by \cite{magain-05}, is likely to be misleading. 
Here we adopt what we consider to be a more robust approach, 
and estimate a virial mass based on the velocity dispersion
($v$) and radius ($R_{BLR}$) of the broad-line region: 
$M_{BH} \sim v^2R_{BLR}/G$ \citep{wandel-99,kaspi-00,vester-02}.

In this method, the BLR velocity dispersion is derived from 
the broad-line widths while the BLR radius is inferred from 
the radius-luminosity relation derived from reverberation mapping 
\citep{kaspi-00,peterson-04}.
Here, we use the recent revision of \cite{kaspi-00}'s virial 
formula presented by \cite{greene-05}. 
This requires measurements of the H$\beta$ FWHM ($v = \sqrt(3)/2 \times$ FWHM)
and the continuum luminosity at 5100\AA ($\lambda L_{5100}$). 
Our measurements of these quantities yield FWHM(H$\beta$) 
$\approx 1270$ km\,s$^{-1}$ and 
$\lambda L_{5100}\approx 4.6\times10^{45}$\,erg\,s$^{-1}$, respectively. 
Inserting these values into equation (5) of Greene \& Ho, we 
obtain $M_{BH} = (9\pm 1)\times 10^7\,M_\odot$. 
Greene \& Ho's alternative virial formula, which employs the 
luminosity and FWHM of the broad H$\alpha$ line yields a consistent 
result, albeit with greater uncertainty: 
$M_{BH} = (6^{+5}_{-3})\times 10^7\,M_\odot$. 
These masses are subject to a systematic uncertainty 
of a factor $\sim 3$ related to the poorly-known structure, kinematics 
and aspect of the BLR (e.g., \cite{onken-04}). 
Nevertheless, at face value, the virial method yields a 
SBH mass that is an {\em order of magnitude} less than 
the value $M_\bullet\approx 8\times 10^8M_\odot$ 
adopted by \cite{magain-05}.

The host galaxies of NLS1's are spirals, often barred \citep{crenshaw-03},
but relatively little is known about their systematic properties.
The tight correlation between bulge velocity dispersion and
SBH mass that characterizes quiescent elliptical galaxies and bulges
\citep{msigma} also appears to be valid for the bulges of active
galaxies, including NLS1s \citep{ferrarese-01,botte-05}.
Adopting $M_\bullet=9\times 10^7M_\odot$, we infer a 
bulge velocity dispersion $\sigma\approx 180$ km s$^{-1}$.
Near-IR bulge luminosities also correlate tightly with $M_\bullet$
\citep{marconi-03}; we infer a $K$-band absolute magnitude
for the stars in the bulge of $M_K\approx -23.4$.
Visual bulge magnitudes are more poorly correlated with $M_\bullet$.
Adopting the \cite{ff-05} relation gives an 
absolute blue magnitude $M_B= -18.9\pm 0.5$;
alternately, applying a $B-K$ color correction of $4.0$ to $M_K$
\citep{peletier-96} gives $M_B\approx -19.4$.
Computing $M_B$ directly from $\sigma$ via the \cite{fj-76}
relation gives a similar value.
Even more uncertain is the predicted total (bulge+disk)
luminosity.
Assuming an Sa host \citep{whittle-92}
implies a disk-to-bulge ratio of $\sim 1.5$ \citep{simien-86}
and a total visual magnitude $M_V\approx -21$.
While very uncertain, this estimate is $2.0-2.5$ magnitudes
fainter than \cite{magain-05}'s estimate ($-23.5\le M_V\le -23.0$) 
based on a $\sim 10\times$ larger assumed value of $M_\bullet$,
and consistent with their
conclusion that the host galaxy must be at least
$4-5$ magnitudes fainter than the quasar ($M_V=-25.8$).

\section{Conclusions}

The HE0450-2958 system consists of a ULIRG that 
experienced a major starburst $\sim 10^8$ yr ago, situated at
$\sim 7$ kpc projected separation from a quasar
having the spectral characteristics of a narrow-line
Seyfert 1.
The quasar host, presumably an early-type  spiral galaxy,
is expected to have a bulge luminosity $M_K\approx -23.4$ ($M_V\approx -21$).
Upper limits on the luminosity of  the quasar host
\citep{magain-05} are consistent with the expected
total luminosity for a galaxy of this type.
The SBH that powers the active nucleus appears to be accreting
at a super-Eddington rate, $L/L_E\approx 3$,
similar to the accretion rates inferred in other
NLS1s.
Ejection of the SBH from the ULIRG is unlikely for a number of reasons, 
the strongest of which is the presence of narrow emission line gas at the
same redshift as the quasar nucleus; this gas could not
have been retained if the SBH was ejected from the companion galaxy.
We find no compelling evidence that the quasar in HE0450-2958 is either
a ``naked'' SBH ejected from its host galaxy,
or that it has an anomalously dark host galaxy.

\end{document}